\documentclass[reprint,
superscriptaddress,
preprintnumbers,
 amsmath,amssymb,
 aps,
]{revtex4-2}

\usepackage{graphicx}% Include figure files
\usepackage{dcolumn}% Align table columns on decimal point
\usepackage{bm}% bold math
\usepackage{orcidlink}
\usepackage{polski}
\usepackage[english]{babel}
\usepackage[subsectionbib]{bibunits}
\hypersetup{hidelinks}

\begin{document}

%\defaultbibliography{apssamp}

%\preprint{APS/123-QED}

\title{Accelerated adiabatic passage of a single electron spin qubit in quantum dots}% Force line breaks with \\

\affiliation{SANKEN, Osaka University, 8-1 Mihogaoka, Ibaraki, Osaka 567-0047, Japan}%
\affiliation{Beijing Academy of Quantum Information Sciences, Beijing 100193, China}%
\affiliation{Lehrstuhl f\"ur Angewandte Festk\"orperphysik, Ruhr-Universit\"at Bochum, Universit\"atsstra{\ss}e 150, Geb\"aude NB, D-44780 Bochum, Germany}
\affiliation{Center for Quantum Information and Quantum Biology (QIQB), Osaka University, Osaka 565-0871, Japan}
\affiliation{Center for Spintronics Research Network (CSRN), Graduate School of Engineering Science, Osaka University, Osaka 560-8531, Japan}
\affiliation{Spintronics Research Network Division, OTRI, Osaka University, Osaka 565-0871, Japan}

\author{Xiao-Fei Liu\orcidlink{0000-0003-0604-6811}}
\email{liuxf@baqis.ac.cn}
\affiliation{SANKEN, Osaka University, 8-1 Mihogaoka, Ibaraki, Osaka 567-0047, Japan}%
\affiliation{Beijing Academy of Quantum Information Sciences, Beijing 100193, China}%

\author{Yuta Matsumoto}
\affiliation{SANKEN, Osaka University, 8-1 Mihogaoka, Ibaraki, Osaka 567-0047, Japan}%

\author{Takafumi Fujita\orcidlink{0000-0002-4678-3069}}
\affiliation{SANKEN, Osaka University, 8-1 Mihogaoka, Ibaraki, Osaka 567-0047, Japan}%\\This line break forced with \textbackslash\textbackslash}%

\author{Arne Ludwig\orcidlink{0000-0002-2871-7789}}
\affiliation{Lehrstuhl f\"ur Angewandte Festk\"orperphysik, Ruhr-Universit\"at Bochum, Universit\"atsstra{\ss}e 150, Geb\"aude NB, D-44780 Bochum, Germany}

\author{Andreas D. Wieck\orcidlink{0000-0001-9776-2922}}
\affiliation{Lehrstuhl f\"ur Angewandte Festk\"orperphysik, Ruhr-Universit\"at Bochum, Universit\"atsstra{\ss}e 150, Geb\"aude NB, D-44780 Bochum, Germany}

\author{Akira Oiwa\orcidlink{0000-0001-5599-5824}}
\email{oiwa@sanken.osaka-u.ac.jp}
\affiliation{SANKEN, Osaka University, 8-1 Mihogaoka, Ibaraki, Osaka 567-0047, Japan}%\\This line break forced with \textbackslash\textbackslash}%
\affiliation{Center for Quantum Information and Quantum Biology (QIQB), Osaka University, Osaka 565-0871, Japan}
\affiliation{Center for Spintronics Research Network (CSRN), Graduate School of Engineering Science, Osaka University, Osaka 560-8531, Japan}
\affiliation{Spintronics Research Network Division, OTRI, Osaka University, Osaka 565-0871, Japan}


\begin{abstract}
Adiabatic processes can keep the quantum system in its instantaneous eigenstate, which is robust to noises and dissipation. However, it is limited by sufficiently slow evolution. Here, we experimentally demonstrate the transitionless quantum driving (TLQD) of the shortcuts to adiabaticity in gate-defined semiconductor quantum dots (QDs) to greatly accelerate the conventional adiabatic passage for the first time. For a given efficiency of quantum state transfer, the acceleration can be more than twofold. The dynamic properties also prove that the TLQD can guarantee fast and high-fidelity quantum state transfer. In order to compensate for the diabatic errors caused by dephasing noises, the modified TLQD is proposed and demonstrated in experiment by enlarging the width of the counterdiabatic drivings. The benchmarking shows that the state transfer fidelity of 97.8\% can be achieved. This work will greatly promote researches and applications about quantum simulations and adiabatic quantum computation based on the gate-defined QDs.

\end{abstract}

\maketitle

%\tableofcontents
\begin{bibunit}

\emph{Introduction.---} Gate-defined semiconductor quantum dots (QDs) can electrically control electron and hole states with ultrahigh precision, which is one of the state-of-the-art quantum devices [1, 2]. The spin qubit of QDs is a  promising candidate for fault-tolerant solid-state quantum computing due to its high-fidelity quantum operation [3--6], potential scalability [7--9], and well compatibility with manufacturing technology of semiconductor industry [10]. Recently, two-qubit gate fidelity of more than 99\% has been demonstrated experimentally [11--14], crossing the well-known surface code threshold [15, 16]. Besides, QD systems are becoming emerging platforms for quantum simulations to explore strongly interacting electrons and topological phases in condensed-matter physics, such as the Fermi–Hubbard system [17], Nagaoka ferromagnetism [18], and the Su–Schrieffer–Heeger model [19]. 

In order to achieve the so called ``quantum advantage'' [20], a high-fidelity quantum processor with large enough computational space and programmable qubits is required. Meanwhile, it also needs accurate quantum control and good robustness against noises and dissipation. One possible pathway is to find a feasible quantum control theory that is applicable for the large-scale quantum processor and guarantees high-accuracy quantum operation simultaneously. It is well known that the manipulation of a quantum state using resonant pulses is sensitive to timing and pulse area errors. In contrast, adiabatic passage can always keep some properties of a dynamical quantum system invariant, ideally switching an initial state into the target state, such as the high fidelity adiabatic process demonstrated in $^{31}$P electron qubit of silicon QD system [21]. This can well prevent decoherence from experimental imperfections [22]. Generally, slow enough evolution is necessary to satisfy adiabatic conditions, limiting its applications. To achieve rapid and robust quantum state manipulation, several shortcuts to adiabaticity (STA) schemes are put forward to compensate for the nonadiabatic errors [23--27], for instance the transitionless quantum driving (TLQD) and invariant-based inverse engineering. Some of them have been demonstrated in other quantum systems [28--33]. Besides, STA has significant applications in quantum simulations to greatly suppress diabatic excitations [34].

\begin{figure}
\includegraphics[width=0.49\textwidth]{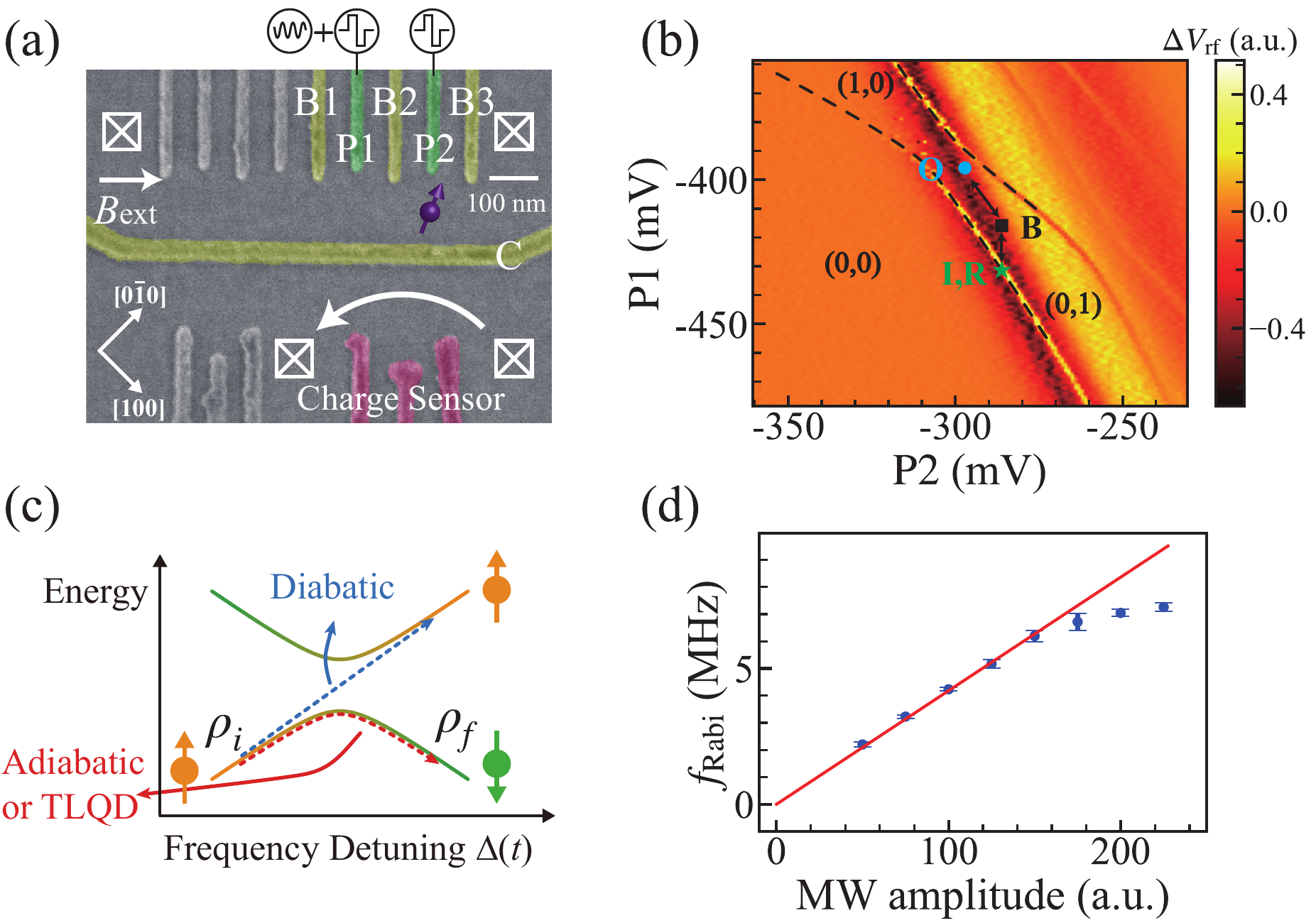}% Here is how to import EPS art
\caption{\label{fig0} The device and its basic properties. (a) The false-colored micrograph of the device. The high frequency pulses are applied through the plunger gates P1 and P2, and the MW driving is connected with P1. (b) Charge stability diagram around single electron region. The position of I, B, and O are represented by the green star, black square, and blue circle, respectively. The position of the initialization is also used for the readout. (c) The schematic of TLQD. (d) Rabi frequency $f_\text{Rabi}$ as a function of the MW amplitude. Its maximum value is about $f_\text{Rabi}^\text{max}\sim7.5$ MHz.}
\end{figure}

Here, we experimentally demonstrate the STA of a single spin qubit in gate-defined QDs for the first time. The experiment is based on the theory of TLQD [23], and the acceleration of quantum state transfer has been achieved. This is also verified from the dynamics of the spin state. To suppress the noises from nuclear spin fluctuations, we propose and experimentally demonstrate a modified TLQD (MOD-TLQD) by enlarging the width of the counterdiabatic pulse. The benchmarking of this MOD-TLQD demonstrates a state transfer efficiency of 97.8\%. Since the gate-defined QDs are moving toward the scalable quantum processor [35], the results of this paper will greatly promote related researches about quantum control and quantum simulations.

\emph{The acceleration of quantum state transfer.---} Figure~\ref{fig0}(a) shows a scanning electron microscope picture of the double QDs (DQDs), which are fabricated on the GaAs/AlGaAs heterostructure. After the implementation of an in-plane magnetic field $B_\text{ext}$, the qubit frequency of a single electron spin is $f_\text{qubit}=\lvert g\rvert\mu_BB/\left(2\pi\hbar\right)$, in which $\mu_B$ is the Bohr magneton, $g$ is the Land{\'e} $g$-factor ($\sim-0.41$ for this GaAs QDs), and $B$ is the total magnetic field (consists of $B_\text{ext}$ and the effective Overhauser field $B^\text{z}_\text{nuc}$). When a microwave (MW) driving is applied, the spin manipulation can be achieved using electric dipole spin resonance [36]. Besides, we use interdot tunneling to enhance the Rabi frequency [37]. We employ energy-selective readout to measure the spin state [38--40]. A nearby charge sensor provides rapid and real time detection of charge state based on the radio frequency (RF) reflectometry [41, 42]. 

Under the rotating frame, the interaction Hamiltonian expanded on the $\lvert\uparrow\rangle$ and $\lvert\downarrow\rangle$ Hilbert space is
\begin{eqnarray} 
\label{eq1}
\hat{H}_0=\frac{\hbar}{2}\left(
\begin{matrix}
-\Delta\left(t\right) & \Omega_R\left(t\right) \\
\Omega_R\left(t\right) & \Delta\left(t\right)
\end{matrix}
\right),
\end{eqnarray}
in which $ \Omega_R\left(t\right)$ is the Rabi frequency, and $\Delta\left(t\right)$ is the frequency detuning with the expression $\Delta\left(t\right)=\omega_\text{qubit}-\omega_\text{MW}-t\dot{\omega}_\text{MW}$. A high-fidelity quantum state transfer can occur if the evolution of this controllable parameter $\Delta\left(t\right)$ is slow enough. However, the TLQD can correct diabatic errors by adding the counterdiabatic driving $\hat{H}_\text{CD}$ even though the evolution does not satisfy adiabatic conditions [23], as shown in Fig.~\ref{fig0}(c). The TLQD can always keep the system in $\lvert\varphi_k\left(t\right)\rangle$, the instantaneous eigenstate of $\hat{H}_0$. Therefore, the time-dependent evolution operator and total Hamiltonian can be obtained. Furthermore, we can know $\hat{H}_\text{CD}$ which has the expression  $i\hbar\sum_k\lvert\partial_t\varphi_k\rangle\langle\varphi_k\rvert$. For this single electron spin system, its specific expression is $\hat{H}_\text{CD}=\hbar\Omega_a\left(t\right)\sigma_y/2$, in which $\Omega_a\left(t\right)=\big[\Omega_R\left( t\right)\dot{\Delta}\left(t\right)-\dot{\Omega}_R\left( t\right)\Delta\left(t\right)\big]/\Omega^2$ and $\Omega^2=\Delta^2\left(t\right)+\Omega_R^2\left(t\right)$. Obviously, the function of $\hat{H}_\text{CD}$ is to correct the diabatic errors by applying a time-dependent driving in $\hat{y}$-axis.

In our experiment, the electron is initialized to $\lvert\uparrow\rangle$ state at the initialization point (I), as shown in Fig.~\ref{fig0}(b). Then, the pulse sequences applied on plunger gates P1 and P2 deliver this electron to the intermediate transit point (B) and then to the operation point (O). After the spin manipulation at O point, this electron is delivered back to B and then to the readout point (R). Here, I and R points are the same. Our setup utilizes an arbitrary waveform generator and an I/Q mixer to precisely tune the time-dependent terms $\Omega_R$, $\Omega_a$, and $\Delta$. The relationship between $\Omega_R$ (or $\Omega_a$) and the MW amplitude has to be characterized firstly. The Rabi frequency estimated from the Rabi oscillation and Landau-Zener transition are nearly the same. Please find more details in Section \uppercase\expandafter{\romannumeral3} of the Supplementary Materials. As shown in Fig.~\ref{fig0}(d), $f_\text{Rabi}$ increases linearly with larger MW amplitude. Then, it becomes saturated progressively until reaching the maximum value $f_\text{Rabi}^\text{max}\sim7.5$ MHz because of the limitation from the trapping potentials or MW amplifiers. 

\begin{figure*}[t]
\includegraphics[width=0.87\textwidth]{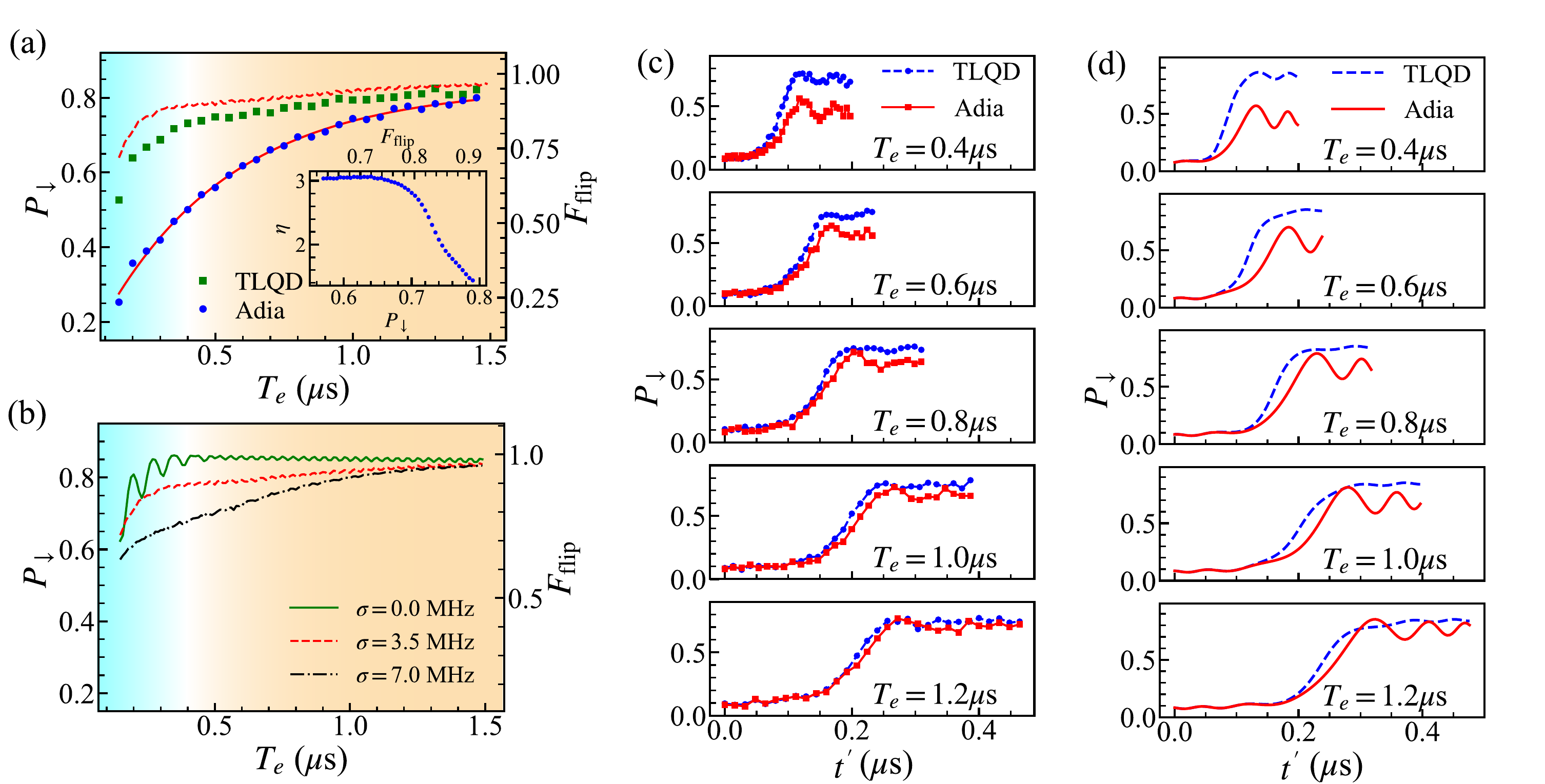}% Here is how to import EPS art
\caption{\label{fig2} The result of TLQD. (a) The final spin down probability $P_\downarrow$ as a function of the evolution time $T_e$ using the conventional adiabatic evolution and TLQD. The red solid line is the fitting to the formula  $AP_\downarrow^\text{LZ}+B$, giving the value of $\Omega_R/2\pi=4.63$ MHz. The inset displays the speedup factor $\eta$ as functions of $P_\downarrow$ and the efficiency of state transfer $F_\text{flip}$. (b) The simulation results of $P_\downarrow$ and $F_\text{flip}$ as a function of $T_e$ under different variance of qubit frequency noise $\sigma\sim$ 0.0 (green solid line), 3.5 (red dashed line), and 7.0 MHz (black dash-dotted line). To better compare the simulation and experimental results, the red dashed line with $\sigma\sim$ 3.5 MHz is also plotted in (a). The modulation depth is $\delta_d=100.0$ MHz. The maximum Rabi frequency is assumed to be $f_\text{Rabi}^\text{max}=7.5$ MHz. (c) and (d) are the experimental and simulation results of the dynamics of $P_\downarrow$, respectively. The Rabi frequency is $\Omega_R/2\pi=4.18$ MHz.}
\end{figure*}

The most significant advantage of this TLQD is that it can always guarantee a quantum system in one of its instantaneous eigenstates and greatly accelerate the adiabatic passage. Figure~\ref{fig2}(a) shows the final spin down probability $P_\downarrow$ and state transfer efficiency (or fidelity) $F_\text{flip}$ as a function of the total evolution time $T_e$. The green squares and blue circles represent the results of TLQD and conventional adiabatic evolution, respectively. The red solid line is the least-squares fitting to the Landau-Zener formula [43--45]. The experimental results show that TLQD always has higher $P_\downarrow$ and $F_\text{flip}$ than the conventional adiabatic passage. The differences of $P_\downarrow$ (also $F_\text{flip}$) between TLQD and adiabatic passage become smaller progressively with longer $T_e$ (slower evolution speed). When $T_e$ is long enough, $\Omega_a$ becomes small enough to be neglected, in analogy to the adiabatic evolution. Note that $F_\text{flip}$ is evaluated from the experimental results $P_\downarrow$ by taking the initialization fidelity ($F_\text{ini}^\uparrow$), spin-to-charge fidelity ($F_\text{STC}^\downarrow$ and $F_\text{STC}^\uparrow$), and charge detection fidelity ($F_\text{E}$) into consideration. Please check Section \uppercase\expandafter{\romannumeral1} and \uppercase\expandafter{\romannumeral6} in the Supplementary Materials. Generally, the relationship $P_\downarrow=P_\downarrow^{\text{ini}=\uparrow}+P_\downarrow^{\text{ini}=\downarrow}$ exists, in which $P_\downarrow^{\text{ini}=\uparrow}$ and $P_\downarrow^{\text{ini}=\downarrow}$ stand for the situations with the initialization of spin to up and down state, respectively. The expressions of $P_\downarrow^{\text{ini}=\uparrow}$ and $P_\downarrow^{\text{ini}=\downarrow}$ are $F_\text{ini}^\uparrow F_\text{flip}F_\text{STC}^\downarrow F_\text{E}+F_\text{ini}^\uparrow \left(1-F_\text{flip}\right)\big(1-F_\text{STC}^\uparrow\big) F_\text{E}$ and $\big(1-F_\text{ini}^\uparrow\big) \big(1-F_\text{flip}\big)F_\text{STC}^\downarrow F_\text{E}+\big(1-F_\text{ini}^\uparrow\big) F_\text{flip}\big(1-F_\text{STC}^\uparrow\big) F_\text{E}$, respectively. We also make sure that the enhancement of state transfer originates from the compensation for diabatic errors instead of simply enlarging the Rabi frequency, please see Section \uppercase\expandafter{\romannumeral2} in the Supplementary Materials. In our experiment, the maximum value of $P_\downarrow$ is about 0.85, which is mainly limited by the readout fidelity. It can be improved by enhancing the relaxation time $T_1$ and bandwidth of the RF-reflectometry after demodulation.

We find that $P_\downarrow$ and $F_\text{flip}$ of TLQD decrease more rapidly when $T_e<$ 0.4 $\mu$s. This originates from the saturation of $\Omega_a$ (because of the large compensation for diabatic errors and the limited value of $f_\text{Rabi}^\text{max}$). Please find the simulation results without considering the limitation of $f_\text{Rabi}^\text{max}$ in Fig.~S13 of the Supplementary Materials. When $T_e>$ 0.4 $\mu$s, there is a tiny increase of $P_\downarrow$ and $F_\text{flip}$. As you can see in Section \uppercase\expandafter{\romannumeral2} of the Supplementary Materials, the TLQD has the highest efficiency of state transfer when $f_\text{qubit}=f_\text{MW}^c$ ($f_\text{MW}^c$ is the center frequency of the MW). The dephasing noises (mainly from the Overhauser field) would cause the fluctuations of $B^\text{z}_\text{nuc}$ and degrade the performance of TLQD. 

\begin{figure}[t]
\includegraphics[width=0.5\textwidth]{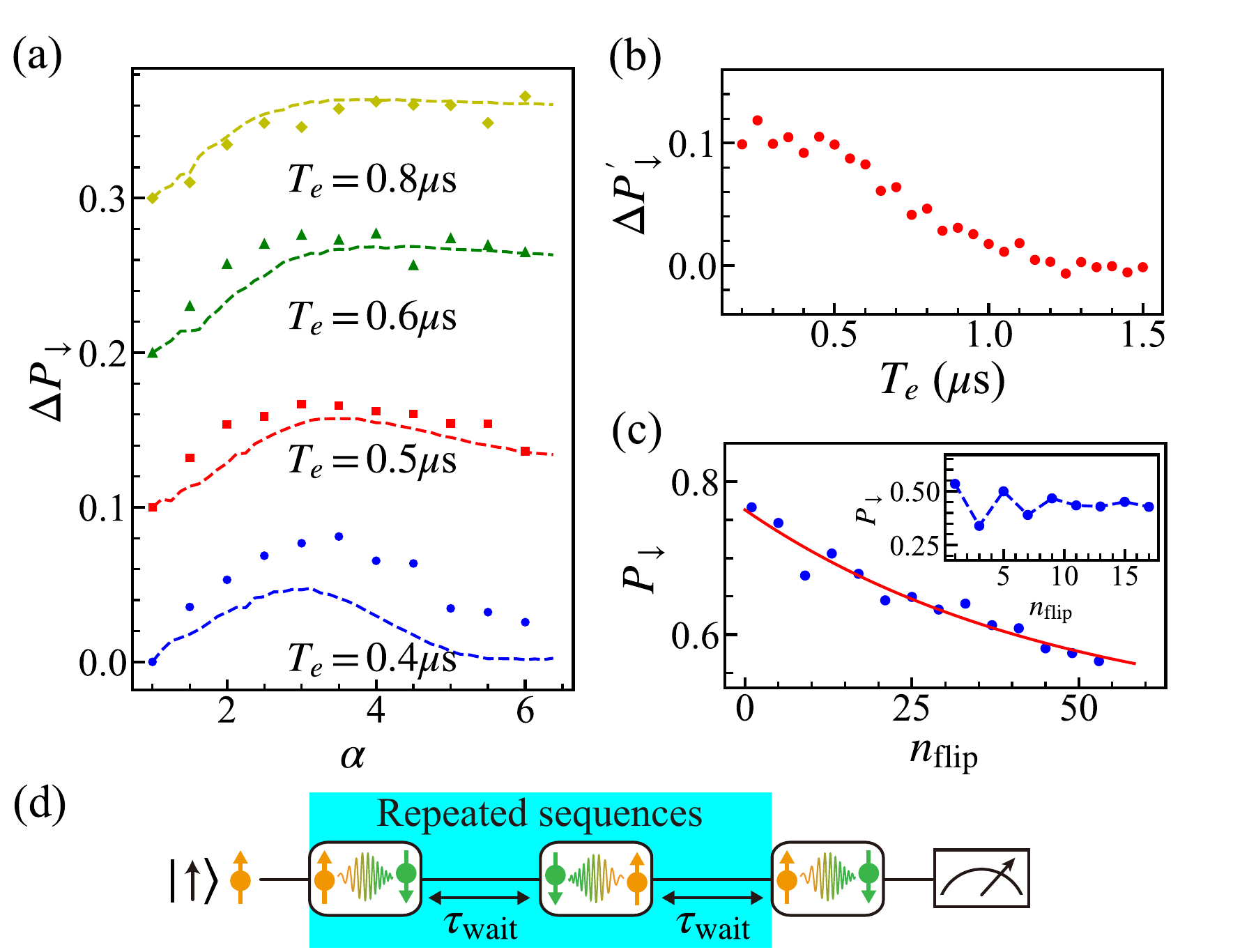}% Here is how to import EPS art
\caption{\label{fig3} The result of MOD-TLQD. (a) The enhancement of spin flip probability $\Delta P_\downarrow$ as a function of $\alpha$ under different $T_e$. The markers are experimental data, and the lines represent simulation results. The variance of qubit frequency is $\sigma\sim$ 2.9 MHz. The traces are shifted vertically for clarity. (b) The enhancement of spin flip probability $\Delta P_\downarrow^\prime$ as a function of $T_e$. The width factor is set to be $\alpha=2.5$. The Rabi frequency in (a) and (b) is $\Omega_R/2\pi \sim 4.0$ MHz. (c) The benchmarking of the efficiency of state transfer using the MOD-TLQD, giving the value of $p=0.978 \pm 0.01$. The inset corresponds to the result of conventional adiabatic evolution, which has the oscillation instead of an exponential decay. (d) The schematic of pulse sequences to benchmark the spin flip fidelity.}
\end{figure}

The simulation after taking dephasing noises and saturation of Rabi frequency into consideration is also performed. For the GaAs QDs [46, 47], the coherence time is dominated by the quasistatic (or low-frequency) noises with a spectral distribution $S\left(f\right)\propto 1/f^\beta$. For simplicity, $\beta$ is set to be 2; i.e., $S\left(f\right)=A^2/f^2$. The variance of the qubit frequency $\sigma$ can be estimated as $\sigma^2=2\int^{1/t}_{f_c}S\left(f\right)df=2A^2\left(1/f_c-t\right)$. Here, $f_c$ and $1/t$ are low and high cutoff frequencies, respectively. The value of $A$ can be calculated from the Ramsey pattern. Using the relationship $1/T_2^*=\sqrt{2}\pi\sigma$, we know $1/T_2^*=2\pi A\sqrt{1/f_c-t}$.  Please find more details in Section \uppercase\expandafter{\romannumeral5} of the Supplementary Materials. Here, the saturation value of total Rabi frequency is $f_\text{Rabi}^\text{max}=7.5$ MHz, i.e., $\Omega\left(t\right)$ is set as 7.5 MHz if $\Omega\left(t\right)>f_\text{Rabi}^\text{max}$. The value of $\sigma$ is about $3.5~\text{MHz}$. The simulation result is plotted as the red dashed line in both Figs.~\ref{fig2}(a) and 2(b), which can well reproduce experimental results qualitatively. For GaAs QDs, $\beta$ may range between 1 and 3. This just changes the value of $A$ without changing the estimation of $\sigma$ too much. In our simulation, we generate 2000 random values of $\delta f_\text{qubit}$ (the shift of the qubit frequency) with the variance $\sigma$. For each $\delta f_\text{qubit}$, we can know $F_\text{flip}$ (also $P_\downarrow$ based on the relationship with $F_\text{flip}$) by solving the Schr\"odinger equation of $\hat{H}_0+\hat{H}_\text{CD}$. The average values of $F_\text{flip}$ and $P_\downarrow$ are the simulation results.

Generally, the TLQD consumes less time compared with conventional adiabatic evolution for a given state transfer efficiency. This acceleration can be characterized quantitatively by the time ratio $\eta=T_\text{adia}/T_\text{TLQD}$, in which $T_\text{TLQD}$ and $T_\text{adia}$ represent the time using the TLQD and conventional adiabatic passage, respectively. The result is shown in the inset of Fig.~\ref{fig2}(a), in which an acceleration of more than twofold can be achieved. The value of $\eta$ becomes flat when $P_\downarrow<0.65$, which is due to the limitation of $f_\text{Rabi}^\text{max}$. Note that $T_\text{TLQD}$ is estimated from the polynomial fitting to the experimental results of TLQD, and $T_\text{adia}$ is deduced from the fitting to the Landau-Zener formula. We believe that the acceleration would be much faster for QDs with longer coherence time, e.g., silicon QDs [48]. The green solid line in Fig.~\ref{fig2}(b) shows the simulation results if $\sigma=$ 0.0 MHz. When the evolution time $T_e>$ 0.4 $\mu$s, $P_\downarrow$ and $F_\text{flip}$ can always keep the highest value. Furthermore, an acceleration of $\eta>6$ can be achieved from our rough estimation. In contrast, large noises would greatly lower the efficiency of state transfer, represented by the black dash-dotted line.

The dynamic properties of TLQD and adiabatic evolution are also investigated experimentally, as shown in Fig.~\ref{fig2}(c). The blue line with circle dots and red line with square dots represent the results of TLQD and conventional adiabatic evolution, respectively. Here, we just show the results starting from the time $0.3~T_e$, i.e., the relative time $t^\prime$ has a shift of $0.3\ T_e$ with respect to the real time. Simulation results are displayed in Fig.~\ref{fig2}(d), which can well reproduce experimental results. The experimental and simulation results show that this TLQD can always keep highest $P_\downarrow$ (also $F_\text{flip}$) after spin flip under various $T_e$ ranging from 0.4 $\mu$s to 1.2 $\mu$s. In contrast, $P_\downarrow$ (also $F_\text{flip}$) would increase gradually with longer $T_e$ for the conventional adiabatic evolution. Meanwhile, its $P_\downarrow$ has much larger amplitude of oscillation compared with TLQD after the spin flip because its quantum state is not the eigenstate of this system. 

\emph{Compensation for dephasing noises.---} For an ideal case, the efficiency of state transfer using TLQD can be up to 100\%. There are two main reasons that make it difficult to realize such high efficiency. The first comes from charge noises, which may cause a shift of the O point and $\Omega_R$, leading to the over- or underestimated value of $\Omega_a$. The second is the nuclear spin fluctuations, which can cause the shift of qubit frequency and significant dephasing in GaAs QDs. Here, we propose a feasible and simple method through pulse optimization to greatly compensate for dephasing noises. 

In the TLQD experiment demonstrated above, $\Omega_R$ is kept as a constant and $\Delta$ is modulated linearly. Therefore, $\Omega_a$ has a Gaussian envelope, i.e., $\Omega_a\left(t\right)\propto\left(\Delta^2+\Omega_R^2\right)^{-1}$. In order to compensate for the dephasing noises, we can enlarge the width of this Gaussian envelope without changing the maximum value of $\Omega_a$. This modified pulse is $\Omega_a^\text{MOD}\left(t\right)=\alpha^2\Omega_R\dot{\Delta}\left(\Delta^2+\alpha^2\Omega_R^2\right)^{-1}$. Here, $\alpha$ is the width factor, and this optimization makes the pulse width to be $\alpha\Omega_R$. The enhancement of $P_\downarrow$, with the definition $\Delta P_\downarrow\left(\alpha\right)=P_\downarrow\left(\alpha\right)-P_\downarrow\left(\alpha=1.0\right)$, as a function of $\alpha$ under various $T_e$ is shown in Fig.~\ref{fig3}(a). It shows that $P_\downarrow$ would increase with $\alpha$ firstly and reach the maximum when $\alpha$ ranges from 2.5 to 3.0. If $T_e<0.6$ $\mu$s, there is a clear drop of $\Delta P_\downarrow$ when $\alpha>2.5$, which may be due to the overcompensation for diabatic errors. In contrast, $\Delta P_\downarrow$ is nearly flat when $\alpha>2.5$ for the situation of $T_e>0.6$ $\mu$s. The reason is that $\Omega_a$ becomes smaller and the effect of overcompensation is not obvious any more. The simulation results shown as the dashed lines can well reproduce our experimental results. We also note that the simulation result of $T_e=0.4$ $\mu$s is much smaller than the experimental result, which may be due to the underestimated value of $f_\text{Rabi}^\text{max}$ in our calculation.

In order to well demonstrate the performance of this width optimization method, the enhancement of $P_\downarrow$ defined as $\Delta P_\downarrow^\prime=\Delta P_\downarrow\left(\alpha=2.5\right)$ as a function of $T_e$ is displayed in Fig.~\ref{fig3}(b). There is a clear enhancement under various $T_e$. Thus, the degradation of state transfer caused by the dephasing noises can be greatly compensated using the MOD-TLQD. Meanwhile, $\Delta P_\downarrow^\prime$ becomes smaller progressively with longer $T_e$ because of the negligible $\Omega_a$. When $T_e>1.1$ $\mu$s, $\Delta P_\downarrow^\prime$ is nearly zero. Besides, the optimal value of $\alpha$ will become smaller with larger $\Omega_R$ because we have to keep $\alpha\Omega_R$ comparable with the dephasing noises. Please see more data in Section \uppercase\expandafter{\romannumeral8} of the Supplementary Materials. 

Finally, the performance of this MOD-TLQD is characterized quantitatively. The probability $P_\downarrow$ as a function of the spin flip number $n_\text{flip}$ is measured, as shown in Fig.~\ref{fig3}(c). The evolution time is $T_e=0.6~\mu\text{s}$, and a waiting time $\tau_\text{wait}=0.2\ \mu\text{s}$ is added after each spin flip process to reduce the thermal heating, as shown in Fig.~\ref{fig3}(d). The repeated sequences represent two flips in a row to keep the spin up state. After fitting to the formula  $P_\downarrow=Ap^{n_\text{flip}}+B$, the fidelity $p=0.978\pm 0.01$ is obtained. The relationship between $n_\text{flip}$ and the number of this repeated sequences $n_\text{seq}$ is $n_\text{flip}=2n_\text{seq}+1$. In contrast, the conventional adiabatic evolution has a clear oscillation for $T_e=0.6~\mu\text{s}$, as shown in the inset in Fig.~\ref{fig3}(c). Only when $T_e$ is large enough (larger than 1.1 $\mu$s), the exponential decay can be observed. More data can be found in Fig.~S12 of the Supplementary Materials. If we perform the spin flip using Rabi oscillation under the same conditions with Fig.~\ref{fig3}(a), i.e., $\Omega_R/2\pi=4.0$ MHz and $\sigma=2.9$ MHz, the spin flip fidelity is less than 65.6\%. Therefore, MOD-TLQD has higher fidelity, although it takes longer time. 

\emph{Conclusion and outlook.---} The STA is experimentally demonstrated in gate-defined QDs for the first time based on the TLQD protocol. Furthermore, the optimization by enlarging the width of counterdiabatic driving can achieve the efficiency of state transfer as high as 97.8\%. The acceleration of quantum state transfer would be much better in Si or Ge QDs with longer coherence time. We also find that the experimental method in our paper can be directly used in the invariant-based inverse engineering [25], which also needs the precise control of time-dependent terms $\Delta\left(t\right)$, $\Omega_R\left(t\right)$, and $\Omega_a\left(t\right)$. Besides, for the cases that the input is a superposition state, i.e., $\left(\lvert\uparrow\rangle+\lvert\downarrow\rangle\right)/\sqrt{2}$, the output state would become $\left(\lvert\uparrow\rangle-\lvert\downarrow\rangle\right)/\sqrt{2}$. It means a $\pi$ rotation along the $\hat{z}$-axis for this superposition state. Meanwhile, the TLQD may be used in other single-qubit operations and adiabatic passages of the QDs system. However, it still needs more researches in both theory and experiment.

\vspace{0.8em}

This work is supported by JST CREST Grant No. JPMJCR15N2; JST Moonshot R\&D Grant No. JPMJMS2066-31 and JPMJMS226B; QSP-013 from NRC, Canada; and the Dynamic Alliance for Open Innovation Bridging Human, Environment and Materials. A. L. and A. D. W. acknowledge the support of DFG-TRR160 and BMBF-QR.X Project 16KISQ009.

%The STA based on the transitionless driving has been demonstrated in gate-defined QD system for the first time. Our calculation by taking the nuclear spin into consideration can well explain the experimental results. In order to improve the fidelity, the pulse shape is optimized. 

\nocite{*}


%\putbib[apssamp]

\end{bibunit}

%\myexternaldocument[prefix]{supplementary_material}
%\onecolumngrid
%\input{supplementary_material.tex}

\begin{bibunit}

\clearpage
\onecolumngrid
%\begin{center}
%    \textbf{\large Supplementary materials for ``Title''}\\[.2cm]
%    Author1$^{1}$ and  Author2$^{1}$ \\[.1cm]
%    {\itshape ${}^1$ Address}
%\end{center}

\begin{center}
    \textbf{\LARGE Supplementary materials}
\end{center}
%---------------------------------------------------------------------------
\maketitle
\setcounter{equation}{0}
\setcounter{section}{0}
\setcounter{figure}{0}
\setcounter{table}{0}
\setcounter{page}{1}
\renewcommand{\theequation}{S\arabic{equation}}
\renewcommand{\thesection}{ \Roman{section}}

\renewcommand{\thefigure}{S\arabic{figure}}
\renewcommand{\thetable}{\arabic{table}}
\renewcommand{\tablename}{Supplementary Table}

\renewcommand{\bibnumfmt}[1]{[S#1]}
\renewcommand{\citenumfont}[1]{#1}
\makeatletter

\maketitle

\setcounter{equation}{0}
\setcounter{section}{0}
\setcounter{figure}{0}
\setcounter{table}{0}
\setcounter{page}{1}
\renewcommand{\theequation}{S-\arabic{equation}}
\renewcommand{\thesection}{ \Roman{section}}

\renewcommand{\thefigure}{S\arabic{figure}}
\renewcommand{\thetable}{\arabic{table}}
\renewcommand{\tablename}{Supplementary Table}

\renewcommand{\bibnumfmt}[1]{[S#1]}
\makeatletter

\maketitle

\section{The device and experimental setup}

\begin{figure}[!h]
\includegraphics[width=0.93\textwidth]{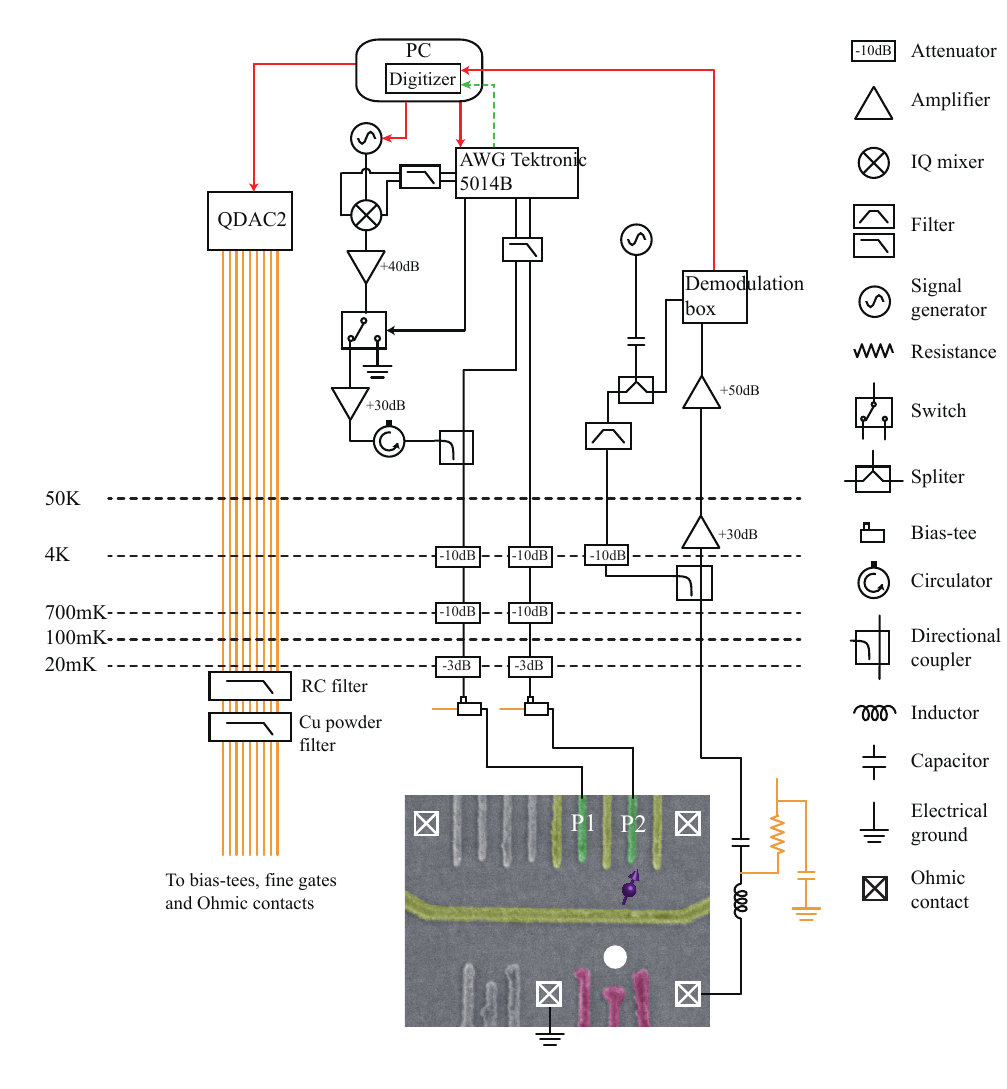}% Here is how to import EPS art
\caption{\label{figs2}The schematic of the setup. All the measurements are performed in the Bluefors dilution refrigerator with the base temperature about 20 mK. The RF-reflectometry technique is used to read out quantum charge state. The MW pulse applied on one plunger gate is modulated by an I/Q mixer.}
\end{figure}

The SEM picture of DQDs is shown in Fig.~1(a) of the main text. The Ti/Au fine electrodes are deposited on the surface of a GaAs heterostructure with 100 nm deep two-dimensional electron gas to apply electrostatic voltages and precisely tune the potentials of QDs. The bias-tees are connected with plunger gates P1 and P2, allowing for the application of MW and nano-second scale pulses. The barrier gates B1, B2, and B3 can be used to control the tunneling strength. An in-plane magnetic field $B_\text{ext}=3.1$ T is applied, corresponding to the resonant frequency of single electron spin qubit $f_\text{qubit}\sim$ 18.0 GHz. The RF-reflectometry provides rapid and real time detection of charge state. All the characterization and measurements are performed in a dilution refrigerator with the electron temperature around 140 mK. Note that there is also a cobalt micromagnet deposited on the surface of this device. However, we do not observe significant enhancement of the Rabi frequency for the rightmost DQDs used in this work. 

The stability diagram of this DQDs system around the single electron charge configuration is shown in Fig.~1(b) of the main text. The number $\left(n_\text{L}, n_\text{R}\right)$ stands for the electron occupation of the left and right QD. The electron is initialized to the spin $\lvert\uparrow \rangle$ state at the position (I) based on the energy-selective tunneling. The tunneling in time $T_\text{in}^\uparrow$ ($T_\text{in}^\downarrow$) of loading a spin $\lvert\uparrow\rangle$ ($\lvert\downarrow\rangle$) electron from the reservoirs is tuned around 2.64 $\mu$s (14.1 $\mu$s), along with the initialization fidelity estimated to be $F_\text{ini}^\uparrow\sim$ 98.8\%. Then, plunger gates P1 and P2 provide nano-second scale pulse sequences to delivery this electron to an  intermediate transit point (B) and then to the operation point (O). The EDSR is implemented by the application of the MW pulse, which is generated by the Keysight N5173B signal generator and I/Q modulated by the signals from the AWG Tektronix 5014B. The MW pulse is applied on the plunger gate P1 through one bias-tee and directional coupler. We employ energy-selective readout to measure the spin state at the readout (R) point, the same with the I point. Note that a detuning $\Delta\epsilon\sim-0.26$ $E_z$ ($E_z$ is the Zeeman splitting) between the Fermi level and the center energy level of $\lvert\downarrow\rangle$ and $\lvert\uparrow\rangle$ states exists at the I and R position. At the readout position R, the tunneling out time $T_\text{out}^\downarrow$ ($T_\text{out}^\uparrow$) of the electron with spin $\lvert\downarrow\rangle$ ($\lvert\uparrow\rangle$) into  reservoirs  is tuned around 3.22 $\mu$s (287.94 $\mu$s). The spin to charge fidelity of $\lvert\downarrow\rangle$ ($\lvert\uparrow\rangle$) state is estimated to be $F_\text{STC}^\downarrow\sim$ 96.7\% ($F_\text{STC}^\uparrow\sim$ 93.9\%). The electrical detection fidelity limited by the bandwidth of the RF-reflectrometry is $F_\text{E}\sim$ 90.4\% for both $\lvert\downarrow\rangle$ and $\lvert\uparrow\rangle$ states [S1].

\newpage
\section{Quantum state transfer under TLQD, adiabatic evolution, and anti-TLQD}
\begin{figure}[!h]
\includegraphics[width=0.99\textwidth]{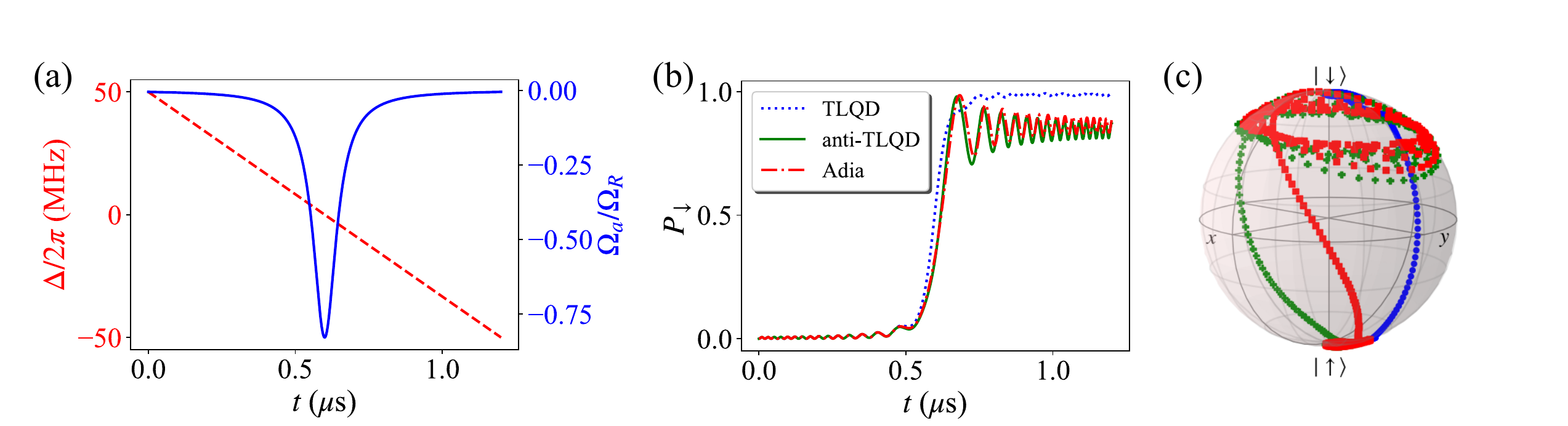}% Here is how to import EPS art
\caption{\label{figs1} (a) Time dependent detuning $\Delta\left(t\right)$ and counter-diabatic driving $\Omega_a\left(t\right)$ when $\Omega_R/2\pi$ is kept as 4.0 MHz. The detuning $\Delta$ is modulated linearly, and $\Omega_a$ follows the Gaussian envelope based on its expression of previous results [S2]. (b) The spin down probability $P_\downarrow$ as a function of the evolution time $t$. The blue dotted line, red solid line, and green dashdotted line correspond to TLQD ($\phi=0$), adiabatic evolution ($\phi=\pi/2$), and anti-TLQD ($\phi=\pi$) situations, respectively. The TLQD can always keep the electron spin in its instantaneous eigenstate state. (c) The evolution of electron spin state in Bloch sphere. The blue circles, red squares, and green pluses represent TLQD, adiabatic evolution, and anti-TLQD, respectively. 
}
\end{figure}

\begin{figure}[!h]
\includegraphics[width=0.7\textwidth]{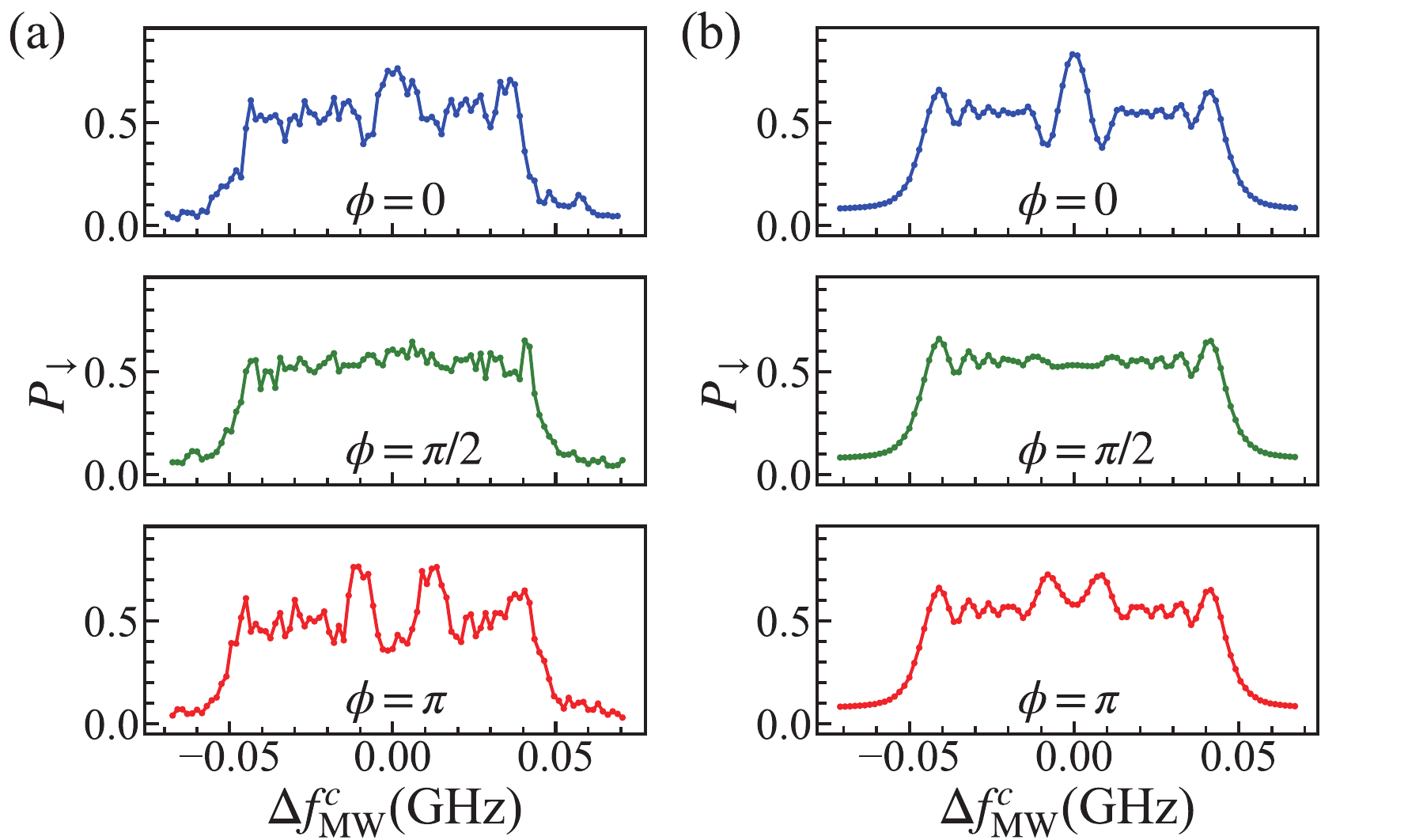}% Here is how to import EPS art
\caption{\label{fig1} The experimental (a) and simulation (b) results of the spin down probability $P_\downarrow$ as a function of the detuning $\Delta f_\text{MW}^c$. The Rabi frequency is $\Omega_R/2\pi=3.44$ MHz. The modulation depth is $\delta_d=100$ MHz, and the evolution time is $T_e=0.6\ \mu$s.}
\end{figure}

To better understand the basic principles and functions of the counter-diabatic term $\Omega_a$, a phase $\phi$ is included, and the corresponding Hamiltonian becomes $\hat{H}_\text{CD}=\hbar\Omega_a\cos\phi\sigma_y/2$. The parameters $\phi=0$, $\phi=\pi/2$, and $\phi=\pi$ correspond to the situations of TLQD, conventional adiabatic evolution, and anti-TLQD, respectively. The anti-TLQD means there is a $\pi$ phase for this counter-diabatic term, i.e., $\Omega_a\rightarrow -\Omega_a$. The experimental data and simulation results of spin down probability $P_\downarrow$ as a function of the detuning $\Delta f_\text{MW}^c$ are shown in Fig.~\ref{fig1}. In this figure, the MW frequency $f_\text{MW}$ is linearly modulated from $f_\text{MW}^c-\delta_d/2$ to $f_\text{MW}^c+\delta_d/2$, in which $f_\text{MW}^c$ is the center frequency of the MW. The detuning $\Delta f_\text{MW}^c=f_\text{qubit}-f_\text{MW}^c$ is the frequency difference between $f_\text{qubit}$ and $f_\text{MW}^c$. For the TLQD ($\phi=0$), $P_\downarrow$ can reach the maximum value when $\Delta f_\text{MW}^c=0$. In contrast, there is always a dip for the anti-TLQD ($\phi=\pi$), while it is flat for the adiabatic evolution ($\phi=\pi/2$). The results mean that the enhancement of state transfer originates from the compensation for diabatic errors instead of simply enlarging the Rabi frequency.

\newpage

\section{Rabi oscillation and Landau-Zener transition}

\begin{figure}[!h]
\includegraphics[width=0.8\textwidth]{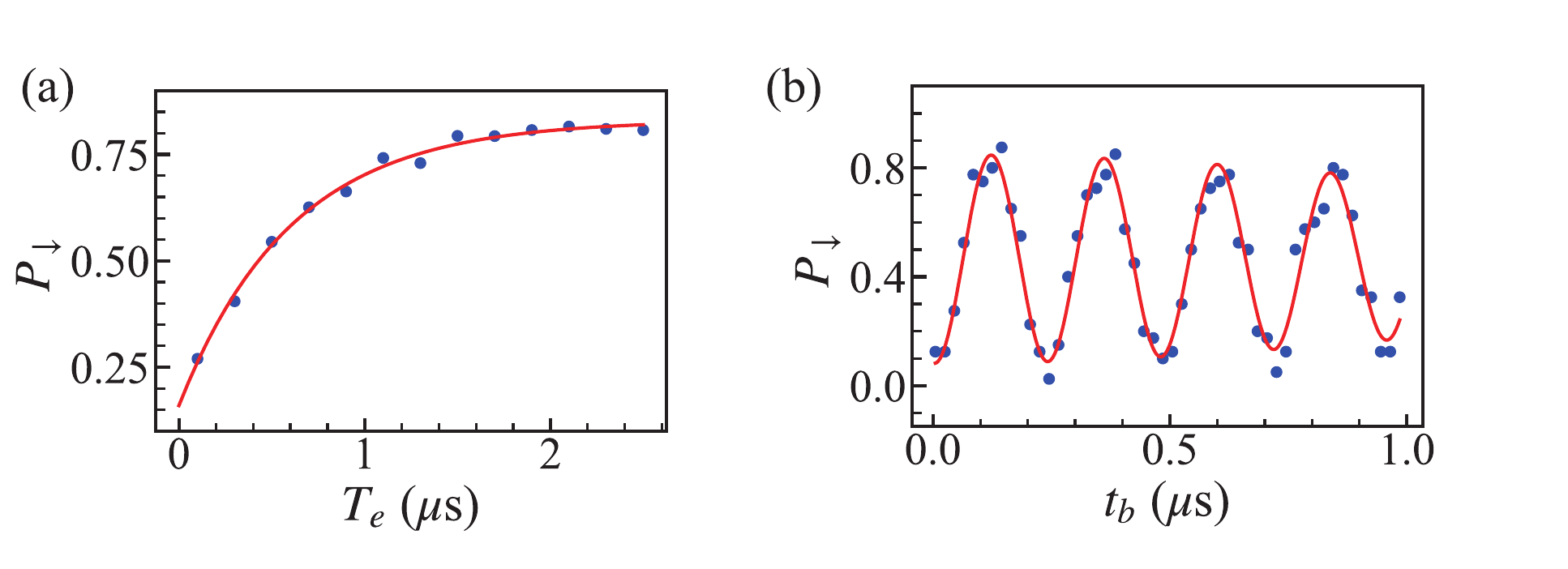}% Here is how to import EPS art
\caption{\label{figs0} (a) The Landau-Zener transition graphic, i.e., the spin down probability $P_\downarrow$ as a function of the  evolution time $T_e$. The modulation depth $\delta_d$ is 100.0 MHz. The experimental value obtained from the fitting of $AP_\downarrow^\text{LZ}+B$ is $\Omega_R/2\pi\sim4.2$ MHz. (b) Rabi oscillation. The voltage for the I/Q mixer is designed to make the same theoretical $\Omega_R$ value with (a). The red fitting curve gives $\Omega_R/2\pi\sim4.1$ MHz.}
\end{figure}

The coherent Rabi oscillation provides a straight forward method to evaluate $\Omega_R$ (or $\Omega_a$). As shown in Fig.~\ref{figs0}(b), the Rabi frequency is estimated to be $f_\text{Rabi}\left(=\Omega_R/2\pi\right)\sim$ 4.1 MHz after fitting to the formula $P_\downarrow\left(t\right)=A\exp(-t^2/T^2_{2,\text{Rabi}})\cos\left(2\pi f_\text{Rabi}t_b\right)+B$. Another optional method to obtain $f_\text{Rabi}$ is utilizing the Landau-Zener transition. For the implementation, the electron is initialized to spin up $\lvert\uparrow \rangle$ state. Then, a linearly modulated MW pulse with the evolution time $T_e$ and depth $\delta_d=100$ MHz is applied. The probability of state transfer to the $\lvert\downarrow \rangle$ state after this frequency modulation is described by the Landau-Zener formula 
\begin{eqnarray}
\label{eq2}
P_\downarrow^\text{LZ}=\exp\left(-\frac{\pi\Omega_R^2}{2\dot{\Delta}}\right). 
\end{eqnarray}
Here, the frequency modulation speed is $\dot{\Delta}=2\pi\delta_d/T_e$. The value of $\Omega_R$ can be evaluated from the fitting to the above equation under various modulation time $T_e$. Figure~\ref{figs0}(a) shows the exponential changes of $P_\downarrow^\text{LZ}$, giving the estimated value of $f_\text{Rabi}\sim4.2$ MHz. The value of $f_\text{Rabi}$ evaluated from the coherent Rabi oscillation and Landau-Zener transition are nearly the same.

\newpage

\section{Pulse sequences}

\begin{figure}[!h]
\includegraphics[width=0.6\textwidth]{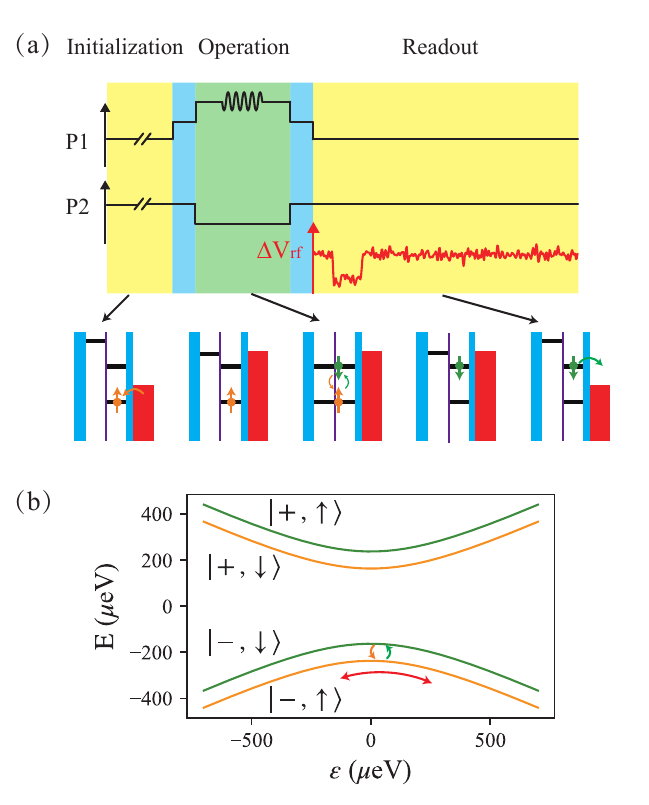}% Here is how to import EPS art
\caption{\label{figs3} (a) The schematic of pulse sequences. The pulses are generated from the AWG Tektronix 5014B, then they are applied on the gates P1 and P2 through the high-frequency cables and bias-tees. The energy selective tunneling is used to readout the charge state [S3]. The red line of $\Delta \text{V}_\text{rf}$ is the trace of signals from the RF-reflectometry when the event of electron tunneling occurs during the readout. (b) The spectrum of DQDs around the zero energy detuning region. The single electron is shuttled between the left and right QD after applying the MW, so there is a quantum state transition between the $\vert -,\uparrow\rangle$ and $\vert -,\downarrow\rangle$ state.}
\end{figure}
%\newpage
%\section{Relaxation rate}
%basic properties of this system, including the T1 time at the operation point and measurement point

In order to enhance the Rabi frequency, the operation point is chosen around the zero detuning region between the left and right QD [S4]. The Hamiltonian of this system after applying an in-plane magnetic field can be written as 
\begin{eqnarray}\label{eq000}
\hat{H}=\frac{\epsilon}{2}\tau_z+t\tau_x+\frac{E_z}{2}\sigma_z,
\end{eqnarray}
in which $\epsilon$ and $t$ are the energy detuning and tunnel coupling between the left and right QD, respectively. $\tau_z$ and $\tau_x$ are Pauli matrixes expanded in the basis of $\vert L\rangle$ anf $\vert R\rangle$, whose expressions are $\tau_z=\vert L\rangle\langle L\vert-\vert R\rangle\langle R\vert$ and $\tau_x=\vert L\rangle\langle R\vert+\vert R\rangle\langle L\vert$. The last term of the Hamiltonian describes the spin term, in which $\sigma_z=\vert\downarrow\rangle\langle\uparrow\vert-\vert\uparrow\rangle\langle\downarrow\vert$ and $E_z$ is the Zeeman splitting. 

Figure~\ref{figs3}(b) shows the spectrum of the Hamiltonian in Eq.~(\ref{eq000}). In this experiment, the electron is shuttled between the left and right QD adiabatically to increase the spin-orbit interaction and Rabi frequency. The value of tunnel coupling $t$ is larger than 200 $\mu$eV and the Zeeman splitting $E_z$ is about 74 $\mu$eV. The adiabatic shuttling between left and right QD is determined by Landau-Zener formula
\begin{eqnarray}
P^\text{LZ}_\text{DQDs}=\exp\left({\frac{-2\pi t^2}{\hbar \dot{\epsilon}}}\right).
\end{eqnarray}
Here, $\dot{\epsilon}\sim 2f_\text{MW}\delta\epsilon$. $\delta\epsilon$ is the change of the detuning and its value is less than 1000 $\mu$eV according to our rough estimation. Therefore, we can know that $P^\text{LZ}_\text{DQDs}\ll1$ and the adiabatic condition can be satisfied.

\newpage

\section{Noise spectrum}

\begin{figure}[!h]
\includegraphics[width=0.6\textwidth]{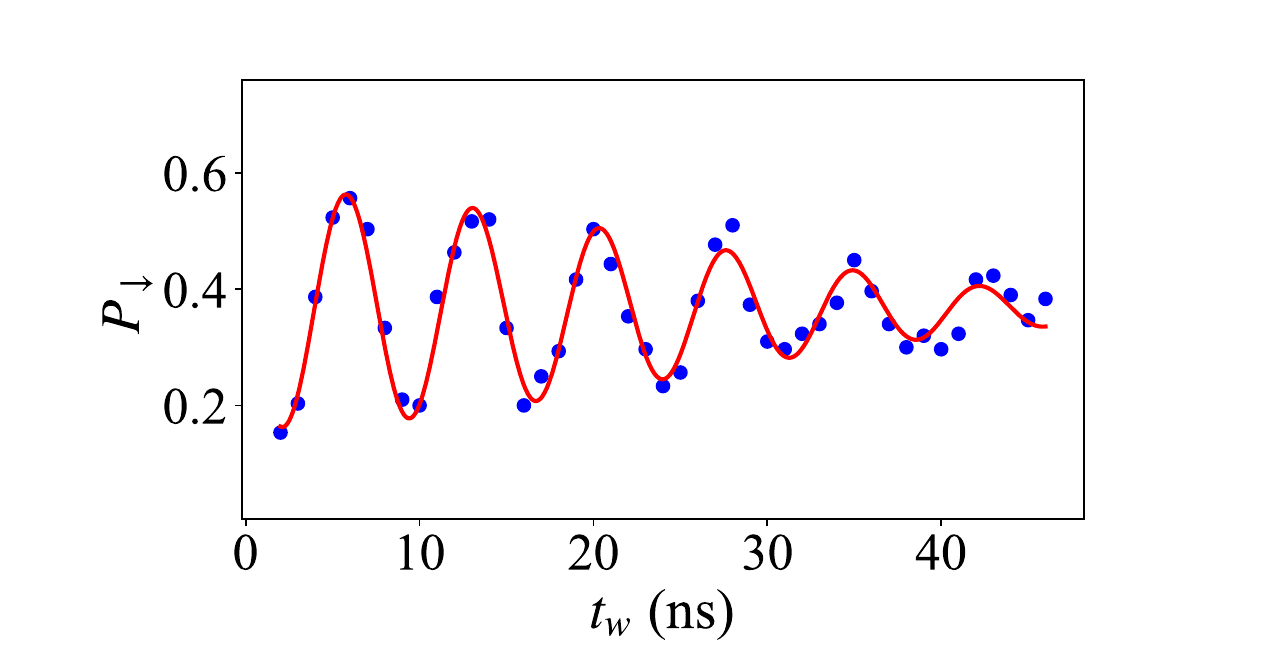}% Here is how to import EPS art
\caption{\label{fig4} Ramsey oscillation. The red line is the fitting curve to the formula $P_\downarrow\left(t\right)=A+B\sin\left(2\pi f t_w+\phi\right)\exp\left[-\left(t_w/T_2^*\right)^2\right]$. The parameters after fitting are $f=136.89$ MHz and $T_2^*=33.2$ ns.}
\end{figure}

The noise spectrum can be extracted from the free-induction decay (FID). Generally, the decay envelope of the phase is dominated by the long corrected noises [S5, S6]. It is often nonexponential and can be characterized by the factor $W\left(t\right)$. For example, $W\left(t\right)$ describing the dephasing caused by the Gaussian noise is
\begin{eqnarray}
W\left(t\right)=\exp\left(-\frac{t^2}{2}\left(2\pi\right)^2\int_{-\infty}^\infty df S_L\left(f\right)\text{sinc}^2\left(\pi ft\right)\right). 
\end{eqnarray}
Here, $\text{sinc}\left(x\right)=\sin x/x$. Since the longitudinal relaxation time is much longer than the pure transverse relaxation time, the dephasing caused by longitudinal relaxation can be neglected. The envelope decay is mainly dominated by the quasistatic noise, i.e., $\lvert f\rvert<1/t$. Meanwhile, $\text{sinc}\left(\pi ft\right)\approx1$ if $\lvert f \rvert\ll 1/t$.  Therefore, this envelope follows the Gaussian decay with the expression as 
\begin{eqnarray}
E_\text{free}\left(t\right)=\exp\left[-\left(\frac{t}{T_2^*}\right)^2\right]=\exp\left[-\frac{t^2}{2}\left(2\pi\sigma\right)^2\right].
\end{eqnarray}
Here, $\sigma=2\int_{f_c}^{1/t}S_L\left(t\right)df$, in which $f_c$ is the lower cut-off frequency. The relationship between the $T_2^*$ and the variance of the qubit frequency is  $T_2^*=1/\left(\sqrt{2}\pi\sigma\right)$. For the GaAs semiconductor QDs, the quasistatic noise is dominated by the Overhauser field, and the two-sided power spectrum is $S\left(f\right)=A^2/f^2$. The coherence time $T_2^*$ extracted from the FID (or Ramsey oscillation) is 
\begin{eqnarray}
1/T_2^*=2\pi A\sqrt{1/f_c-t}.
\end{eqnarray}
 
Figure~\ref{fig4} shows the Ramsey oscillation with the lower cut-off frequency $f_c\sim0.71$ Hz, giving the value of $A\sim 4.043\times10^6\ \text{Hz}^{\frac{3}{2}}$.

The variance of the qubit frequency $\sigma$ is different in Fig.~2(a) and Fig.~3(a) of the main text, which is mainly due to the different value of $f_c$. In our experiment, we load all the waveform under different $T_e$ (or $\alpha$) to the AWG simultaneously. In Fig.~2(a), each one single-shot measurement under different $T_e$ ranging from 0.15 $\mu$s to 1.45 $\mu$s is performed in serious. This process is repeated until we finish the measurement. Therefore, the value of $1/f_c$ is set to be the total measurement time. We can also know the value of $\sigma$ using the same method for Fig. 3(a) under different $\alpha$.

\newpage 
\section{Initialization and readout fidelity}

\begin{figure}[!h]
\includegraphics[width=0.8\textwidth]{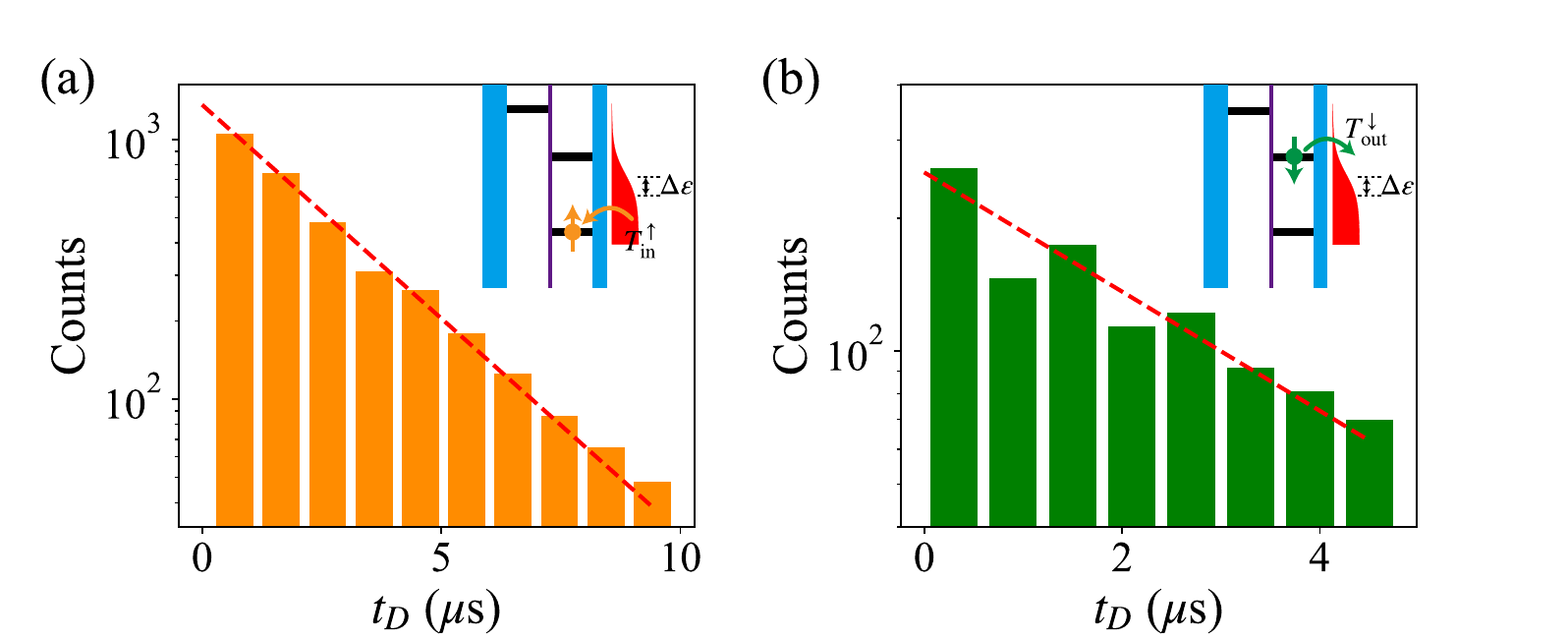}% Here is how to import EPS art
\caption{\label{fig6} Histogram showing the tunneling in time $T_\text{in}^\uparrow$ of $\lvert\uparrow\rangle$ state (a) and tunneling out time out $T_\text{out}^\downarrow$ of $\lvert\downarrow\rangle$ state (b), whose values are 2.64 $\mu$s and 3.22 $\mu$s, respectively.}
\end{figure}

This section will introduce the method used for the analyses of state preparation and measurement (SPAM). The following analyses of fidelity and naming scheme are proposed by D. Keith \emph{et al.}  in the paper ``New Journal of Physics \textbf{21}, 063011 (2019)'' [S1]. Here, we use their method and naming scheme to analyze the properties of our sample. 

The readout fidelity of $\lvert\uparrow\rangle$ and $\lvert\downarrow\rangle$ state are analyzed firstly. The loading time $T_\text{in}^\uparrow$ of the $\lvert\uparrow\rangle$ state and tunneling out time $T_\text{out}^\downarrow$ of the $\lvert\downarrow\rangle$ state can be measured directly, as shown in Fig.~\ref{fig6}. The values of $T_\text{in}^\uparrow$ and $T_\text{out}^\downarrow$ extracted from the histograms are 2.64 $\mu$s and 3.22 $\mu$s, respectively. We find that the Fermi level $E_F$ of the reservoirs is not in the center position of the $\lvert\uparrow\rangle$ and $\lvert\downarrow\rangle$ state, which means that a detuning $\Delta\epsilon$ exists. Its value can be estimated from the following relationship 
\begin{eqnarray}
\frac{T_\text{out}^\downarrow}{T_\text{in}^\uparrow}=\frac{f\left(\Delta\epsilon-E_z/2\right)}{1-f\left(\Delta\epsilon+E_z/2\right)}.
\end{eqnarray}
Here, $f\left(\epsilon\right)$ is the Fermi–Dirac distribution with the expression $f\left(\epsilon\right)=\left[1+\exp\left({\epsilon/k_BT}\right)\right]^{-1}$. $E_z$ is the Zeeman splitting after applying the in-plane magnetic field. $k_B$ is the Boltzmann constant. The electron temperature is $T\sim140 $ mK. The value of this detuning is $\Delta\epsilon\sim-0.26\ E_z$. Furthermore, the tunneling out time $T_\text{out}^\uparrow$ of the $\lvert\uparrow\rangle$ state and the loading time $T_\text{in}^\downarrow$ of the $\lvert\downarrow\rangle$ state are estimated to be 287.94 $\mu$s and 14.12 $\mu$s, respectively. This estimation is based on the relationship
\begin{eqnarray}
\frac{T_\text{in}^\uparrow}{T_\text{in}^\downarrow}=\frac{f\left(\Delta\epsilon+E_z/2\right)}{f\left(\Delta\epsilon-E_z/2\right)},\  \   \text{and}\ \ \    \frac{T_\text{out}^\uparrow}{T_\text{out}^\downarrow}=\frac{1-f\left(\Delta\epsilon+E_z/2\right)}{1-f\left(\Delta\epsilon-E_z/2\right)}. 
\end{eqnarray}

\begin{figure}[!h]
\includegraphics[width=0.8\textwidth]{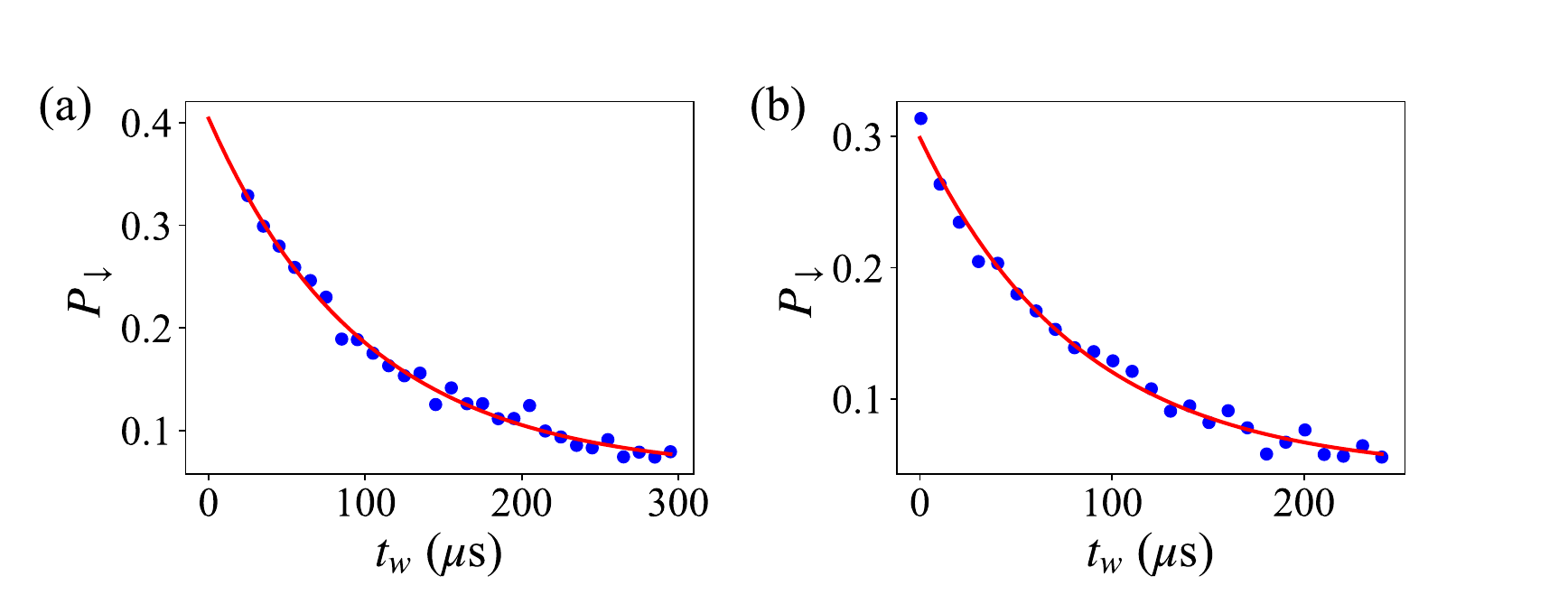}% Here is how to import EPS art
\caption{\label{fig5} The relaxation time $T_1$ of the spin down $\lvert \downarrow\rangle$ state at the R point (a) and O point (b). The fitting values of $T_1$ in (a) and (b) are 99.5 $\mu$s and 83.3 $\mu$s, respectively. }
\end{figure}

The spin-to-charge fidelity can be calculated using the following formula
\begin{eqnarray}
F_\text{STC}^\uparrow&=&e^{-\frac{t}{T_\text{out}^\uparrow}},\nonumber \\
F_\text{STC}^\downarrow&=&\frac{1}{T_\text{out}^2}\left[\left(1-e^{-\frac{t}{T_\text{out}^\uparrow}}\right)T_\text{out}^\uparrow T_\text{out}^\downarrow+\left(e^{-\frac{T_1+T_\text{out}^\downarrow }{T_\text{out}^\downarrow T_1}t}-1\right)T_1  \left(T_\text{out}^\downarrow-T_\text{out}^\uparrow\right) \right]. 
\end{eqnarray}
in which $T_\text{out}^2=T_1\left(T_\text{out}^\uparrow-T_\text{out}^\downarrow\right)+T_\text{out}^\uparrow T_\text{out}^\downarrow$. The relaxation time of the spin down state $\lvert\downarrow \rangle$ is $T_1\sim\ 99.5\ \mu$s, as shown in Fig.~\ref{fig5}(a). The readout time is about 18.0 $\mu$s. Therefore, the values of $F_\text{STC}^\downarrow$ and $F_\text{STC}^\uparrow$ are 96.7\% and 93.9\%, respectively. 

\begin{figure}[!h]
\includegraphics[width=0.5\textwidth]{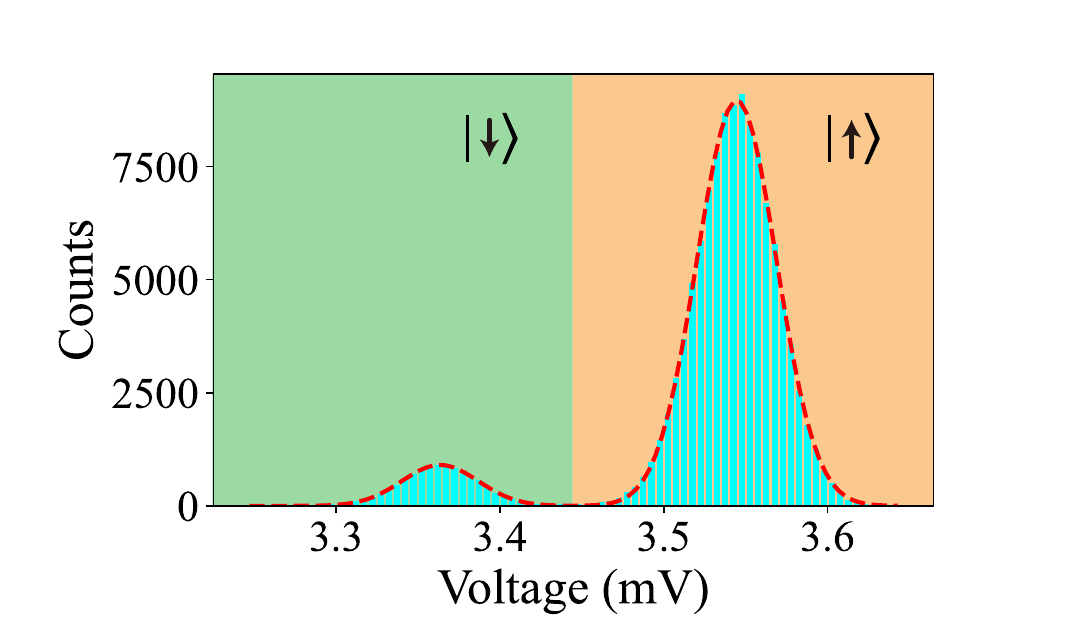}% Here is how to import EPS art
\caption{\label{fig7} Histogram of the demodulated RF signals at the R point. The red dashed line is the double Gaussian fitting. This distribution corresponds to the infidelity of less than 0.01\%, which can be neglected.}
\end{figure}

The total readout fidelity is also determined by the electrical detection fidelity, limited by the signal-to-noise ratio (SNR), bandwidth of the RF-reflectometry, and the sampling rate of the digitizer. Figure~\ref{fig7} shows the histogram of signals from the RF-reflectometry. If the threshold determining the tunneling out event is set at the center position of two Gaussian envelopes.  The infidelity is less than 0.01\%, meaning the SNR is high enough. The corresponding infidelity is neglected in the following analyses. Therefore, the electrical detection fidelity is only determined by the probability of missing the ``fast blip'', which means that one electron tunnels out and another electron tunnels into the QDs quickly within the resolution time of the setup. In our experiment, the sampling rate of the digitizer is 10.0 MHz, and the bandwidth of the RF-reflectometry is 1.9 MHz. This ``fast blip'' occurs within the time $t_s\sim0.53\ \mu$s.

The tunneling out probability density of $\lvert\uparrow\rangle$ and $\lvert\downarrow\rangle$ state within the time $t_\text{s}$ is 
\begin{eqnarray}
p^{\downarrow\left(\uparrow\right)}_\text{out}\left(t\right)=\frac{e^{-t/T_\text{out}^{\downarrow\left(\uparrow\right)}}}{T_\text{out}^{\downarrow\left(\uparrow\right)}(1-e^{-t_\text{s}/T_\text{out}^{\downarrow\left(\uparrow\right)}})},\ \ \ \ \ \  0<t<t_\text{s}.
\end{eqnarray}
The tunneling in probability density of $\lvert\uparrow \rangle$ state after the tunneling out of one electron is 
\begin{eqnarray}
p^\uparrow_\text{in}\left(t\right)=\frac{e^{-t/T_\text{in}^\uparrow}}{T_\text{in}^\uparrow}. 
\end{eqnarray}
Therefore, the probability of missing the detection signal of tunneling out $\lvert\uparrow\rangle$ or $\lvert\downarrow\rangle$ state is
\begin{eqnarray}
P_\text{miss}^{\downarrow\left(\uparrow\right)}\left(t\right)=\int_0^{t_\text{s}}\int_0^t p^{\downarrow\left(\uparrow\right)}_\text{out}\left(t-\tau\right) p^\uparrow_\text{in}\left(\tau\right) d\tau dt. 
\end{eqnarray}
From this calculation, we know the electrical charge detection fidelity of the $\lvert\downarrow\rangle$ state and $\lvert\uparrow\rangle$ state are 90.4\% and  state 90.6\%, respectively.

Finally, we calculate the initialization fidelity of the spin up $\lvert\uparrow \rangle$ state. The rate equation during the initialization stage is $d\left(P_0,P_\uparrow,P_\downarrow\right)^\text{T}/dt=\bm{M}\left(P_0,P_\uparrow,P_\downarrow\right)^\text{T}$. Here, $P_0$, $P_\uparrow$, and $P_\downarrow$ represent the probability of no electron, one electron with spin up, and one electron with spin down in the DQDs, respectively. The expression of $\bm{M}$ is 
\begin{eqnarray} 
\label{eq0}
\bm{M}=\left(
\begin{matrix}
-\dfrac{1}{T_\text{in}^\uparrow}-\dfrac{1}{T_\text{in}^\downarrow}& \dfrac{1}{T_\text{out}^\uparrow} & \dfrac{1}{T_\text{out}^\downarrow}\\
\dfrac{1}{T_\text{in}^\uparrow} & -\dfrac{1}{T_\text{out}^\uparrow} &\dfrac{1}{T_1} \\
\dfrac{1}{T_\text{in}^\downarrow} & 0 &-\dfrac{1}{T_\text{out}^\downarrow}-\dfrac{1}{T_1}
\end{matrix}
\right).
\end{eqnarray}
The initialization time is 30 $\mu$s. If the initial state of the QDs is empty, the initialization fidelity of the spin up $\lvert\uparrow\rangle$ state is $F_\text{ini}^\uparrow=P_\uparrow\left(t=30 \mu\text{s}\right)\sim$ 98.8\%.

\newpage 

\section{The results of TLQD under different $\Omega_R$}
%different gate voltage--> all shows the same enhancement. 

\begin{figure}[!h]
\includegraphics[width=1.05\textwidth]{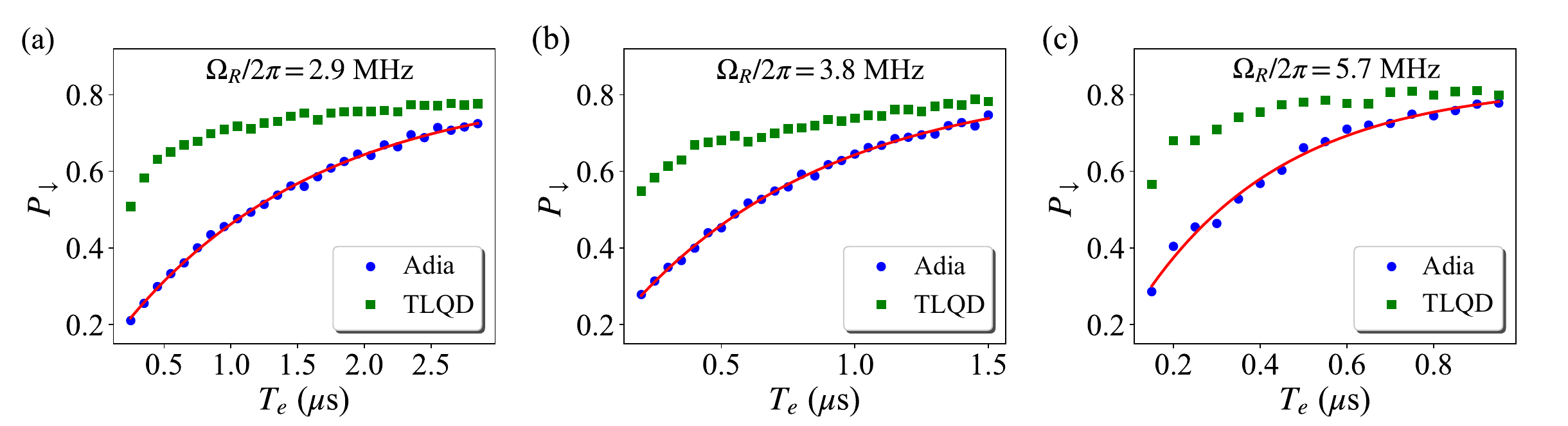}% Here is how to import EPS art
\caption{\label{fig8} The spin down probability $P_\downarrow$ as a function of the total evolution time $T_e$ under different $\Omega_R$. The blue circles and green squares correspond to the adiabatic evolution and TLQD, respectively. The red lines are the fitting results using the Landau-Zener formula. }
\end{figure}

\newpage

\section{Pulse optimization}
%the effect of width--> different evolution time

\begin{figure}[!h]
\includegraphics[width=0.8\textwidth]{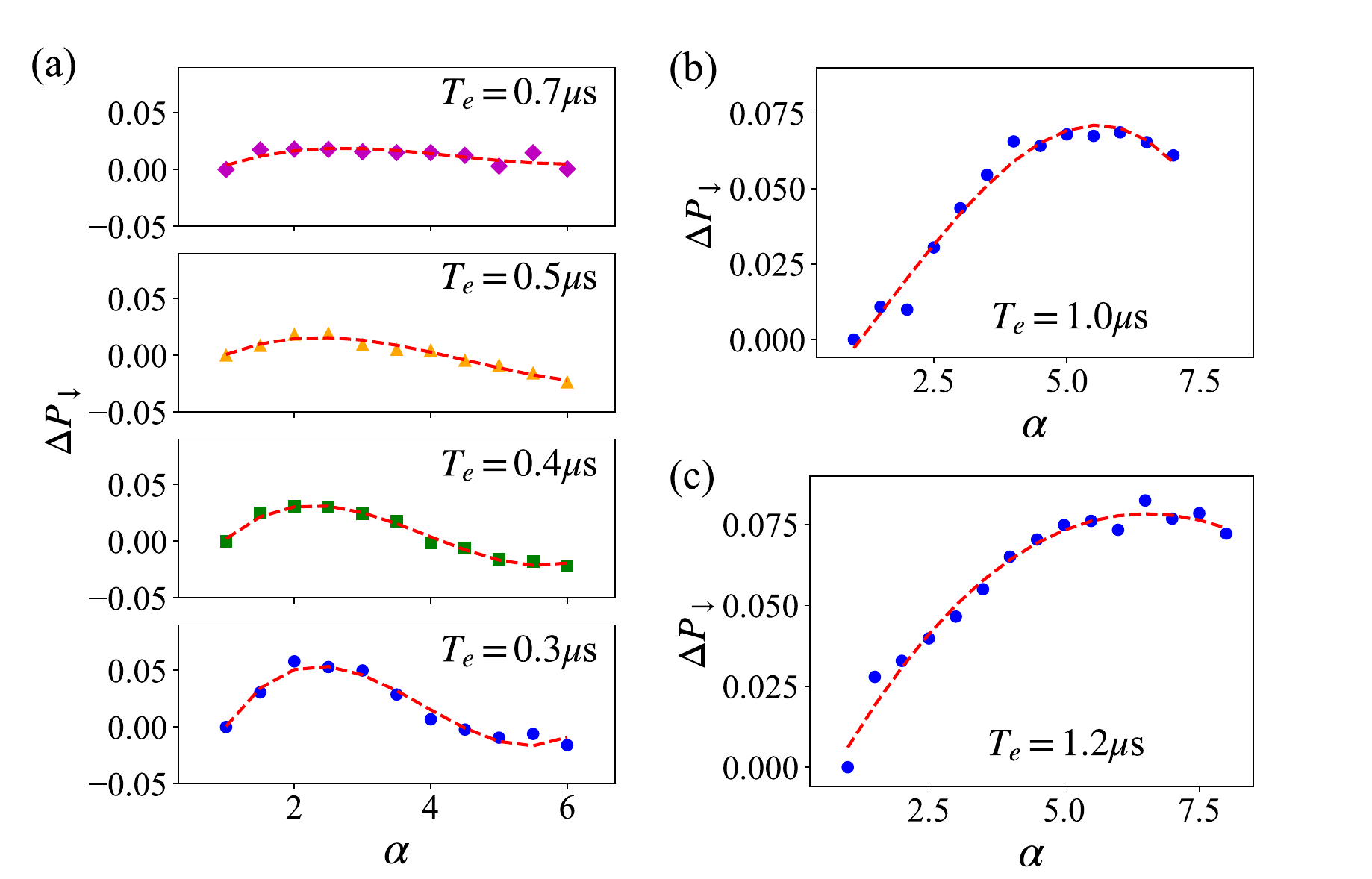}% Here is how to import EPS art
\caption{\label{fig00} The width factor $\alpha$ dependence of $\Delta P_\downarrow$. The values of $\Omega_R$ in (a), (b), and (c) are 4.8 MHz, 2.9 MHz, and 2.4 MHz, respectively. The red dashed lines are the polynomial fitting curves. The optimal width factor $\alpha$ are about 2.0 (a), 5.5 (b), and 6.5 (c), respectively. Generally, this optimal value becomes smaller with larger $\Omega_R$. }
\end{figure}

\newpage 
\section{The benchmarking of spin flip fidelity}
\begin{figure}[!h]
\includegraphics[width=0.8\textwidth]{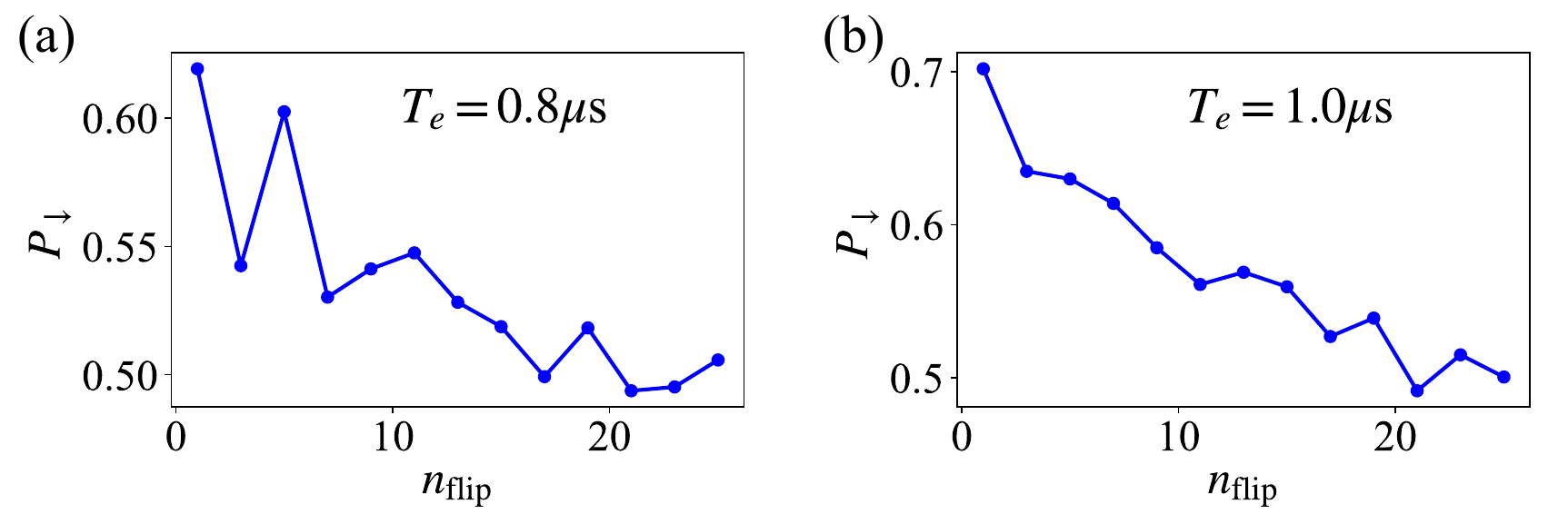}% Here is how to import EPS art
\caption{\label{fig9} (a) and (b) show the spin down probability $P_\downarrow$ as a function of spin flip number $n_\text{flip}$ using the conventional adiabatic evolution when the total evolution time $T_e$ are 0.8 $\mu$s and 1.0 $\mu$s, respectively. When the $T_e$ is short, e.g., 0.8 $\mu$s, $P_\downarrow$ can not show exponential decay, indicating the $F_\text{flip}$ is very small. $P_\downarrow$ start to follow an exponential decay when $T_e$ is long enough. This is completely different from the TLQD theory, which can always exhibit exponential decay, even though $T_e$ is very short.}
\end{figure}

\newpage
\section{The results of TLQD without the saturation of Rabi frequency}
%different gate voltage--> all shows the same enhancement. 

\begin{figure}[!h]
\includegraphics[width=0.5\textwidth]{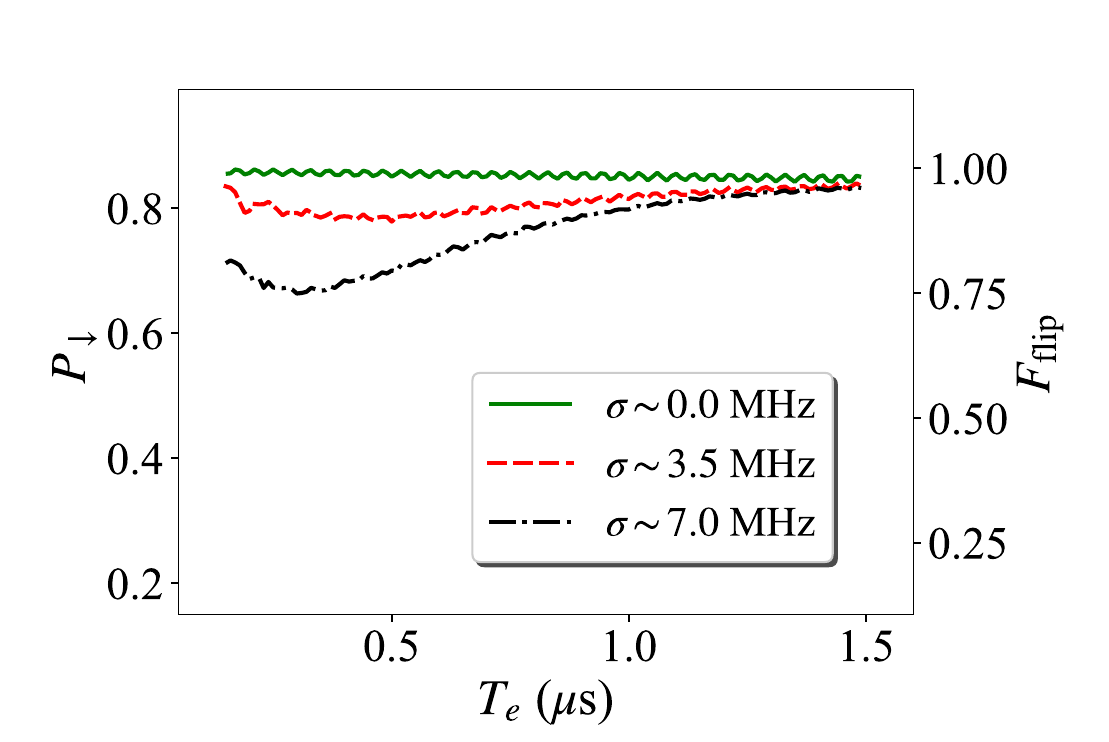}% Here is how to import EPS art
\caption{\label{fig8} The simulation results of the spin down probability $P_\downarrow$ and the state transfer efficiency $F_\text{flip}$ of the TLQD scheme without considering the saturation of Rabi frequency.}
\end{figure}

\vspace{1em}

%We wish to acknowledge the support of the author community in using REV\TeX{}, offering suggestions and encouragement, testing new versions,

\nocite{*}
%\bibliography{aipsamp}% Produces the bibliography via BibTeX.

\end{bibunit}

\end{document}